\documentclass[12 pt]{iopart}

\usepackage{graphicx}
\usepackage{dcolumn}
\usepackage{bm}

\newcommand{\pderiv}[2]{\frac{\partial #1}{\partial #2}}

\begin{document}

\title{Phase transitions in a lattice population model}

\author{Alastair Windus and Henrik Jeldtoft Jensen}
\ead{h.jensen@imperial.ac.uk; www.ma.ic.ac.uk/$\:\tilde{}\:$hjjens}
\address{Department of Mathematics, Imperial College London,
South Kensington Campus, London. SW7 2AZ.}

\begin{abstract}
We introduce a model for a population on a lattice with diffusion and birth/death according to $2A\longrightarrow 3A$ and $A\longrightarrow\phi$ for
a particle $A$. We find that the model displays a phase transition from an
active to an absorbing state which is continuous in $1+1$ dimensions and of
first-order in higher dimensions in agreement with the mean field equation.
For the $1+1$ dimensional case, we examine the critical exponents
and a scaling function for the survival probability and show that it belongs to the universality class of directed percolation. In higher dimensions,
we look at the first-order phase transition by plotting a histogram of the
population density and use the presence of phase coexistence to find an accurate value for the critical point in $2+1$ dimensions. 
\end{abstract}


\section{Introduction} \label{s: Introduction}
Non-equilibrium phase transitions have long been a major area of investigation (see
\cite{Hinrichsen_Non} for a review). Studies are wide-ranging, involving atmospheric precipitation \cite{Peters}, sandpile models \cite{Vespignani},
epidemics \cite{Dammer} and many more. One of the great achievements in the
field has been the discovery that a broad range of models belong to one of
a few universality classes, whose members share the same critical exponents and scaling functions (see for example \cite{Lubeck_Universal}). In particular, it has been conjectured
that all models with a scalar order-parameter that exhibit a continuous phase transition from an active state to a single absorbing state belong to the
same universality class of \textit{directed
percolation} (DP) \cite{Grassberger_On, Janssen_Non}. Although not proven,
the conjecture is strongly supported by numerical evidence and seems
to be even more general since, for example, a system with multiple absorbing states is known to belong to the class \cite{Jensen_Dickman}.

Here, we introduce a lattice model representing a population in a habitat
and include the processes of birth, death and diffusion. Due to the conflict
between growth and decay, with steady state population density
$\bar\rho$ as our order parameter, we expect a phase
transition to an absorbing state to occur under certain conditions. The expected
phase transition may be either continuous or of first-order. The latter, however, is rarely seen in low spatial
dimensions due to the destabilisation of the ordered phase caused by the
larger fluctuations that are present in such systems. Hinrichsen 
has hypothesised that first-order phase transitions are impossible in $1+1$
dimensional systems provided that there are no additional conservation laws,
long-range interactions, macroscopic currents or special boundary conditions
\cite{Hinrichsen_First-order}.
Since our model does not fulfil any of these criteria, we expect at least
the $1+1$ dimensional version of our model to exhibit a continuous phase transition
and thus to belong to DP.

Thanks to series expansions, the critical exponents of DP are now known to
a high degree of accuracy \cite{Jensen_Low}. We therefore proceed in Section
\ref{s: The Model} by describing our model and then, in Sections \ref{s: Critical
Exponents} and \ref{s: Scaling Functions}, examine the critical exponents and a scaling function respectively to compare them with those
of DP. In Section \ref{s: First-order}, we examine the first-order phase transitions and conclude with some remarks in Section \ref{s: Conclusions}.

\section{The Model} \label{s: The Model}
We have a $d$-dimensional square lattice of linear length $L$ where each square is either
occupied by a single particle (1) or is empty (0). A site is chosen at random. With probability $p_{\rm d}$
the particle on an occupied site dies,  leaving the site empty. If the particle does not die, a nearest neighbour site is randomly chosen. If the neighbouring site
is empty the particle moves there; otherwise, the particle reproduces with probability
$p_{\rm b}$ producing a new particle on another randomly selected neighbouring
site, conditional on that square being empty. A time step is defined
as the number of lattice sites $N=L^{\rm d}$ and periodic boundary conditions are
used. 
We have the following reactions for a particle $A$ for proliferation and
annihilation respectively,
        \begin{equation} \label{Reactions}
        A+A\longrightarrow3A \quad \mbox{and} \quad A\longrightarrow \phi.
        \end{equation}
Our model is similar to that of Schl\"ogl's second model \cite{Schlogl} except for the inclusion of diffusion and the reactions given in (\ref{Reactions})
being unidirectional.

Assuming the particles are spaced homogeneously, the mean field equation
for the density of active sites $\rho(t)$ is given by
        \begin{equation}
        \pderiv{\rho(t)}{t} = p_{\rm b}(1-p_{\rm d})\rho(t)^2(1-\rho(t))-p_{\rm         d}\rho(t).
        \end{equation}
This has three stationary states
        \begin{equation} \label{steady states}
        \bar\rho_0 = 0, \quad \bar\rho_\pm=\frac{1}{2}\left(1\pm\sqrt{1-\frac{4p_{\rm         d}}{p_{\rm b}(1-p_{\rm d})}}\right).
        \end{equation}  
Clearly, for $4p_{\rm d} > p_{\rm b}(1-p_{\rm d})$, $\bar\rho_0$ is the only real stationary state,
resulting in a phase transition occurring at the critical death rate $p_{{\rm
d}_{\rm c}}
=p_{\rm b}/(4+p_{\rm b})$. Simple analysis shows that $\bar\rho_+$ and $\bar\rho_0$ are stable stationary states,
whereas $\bar\rho_-$ is unstable and therefore represents a critical density $\rho_{\rm c}$ below which extinction will occur in all cases. So, for $p_{\rm d} < p_{{\rm d}_{\rm c}}$,
        \begin{equation}
        \rho(t) \longrightarrow \quad \left\{ 
        \begin{array}{cl}
        0 &  \mbox{ for } \rho(t) < \rho_{\rm c} \\
        \bar\rho_+ & \mbox{ for } \rho(t) > \rho_{\rm c}
        \end{array} \right.
        \quad\mbox{ as } t \longrightarrow \infty. 
        \end{equation} 

\begin{figure}[tb]
        \centering\noindent
        \includegraphics[width=8cm]{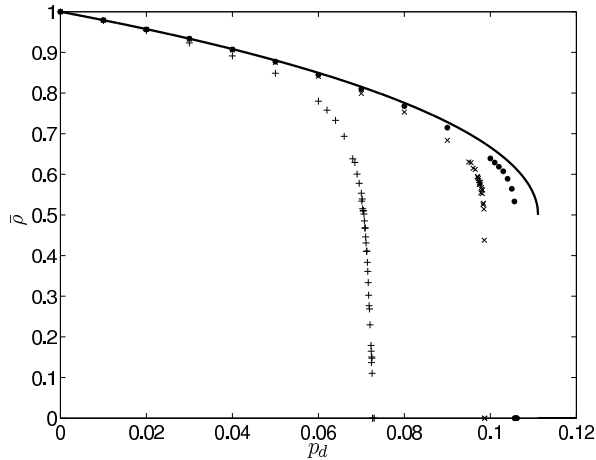}
        \caption{Steady state population densities for the mean field (line)
        and the $1+1$ ($+$), $2+1$ ($\times$) and $3+1$ ($\bullet$) dimensional
        Monte Carlo simulations.}
        \label{SteadyState}
        \end{figure}

At $p_{\rm d} = p_{{\rm d}_{\rm c}}$, the stationary density jumps from 1/2 to 0, resulting
in a first-order phase transition. We investigate whether the Monte Carlo (MC) simulations agree with the mean field by
plotting, in Figure \ref{SteadyState}, 
the steady state population density against $p_{\rm d}$ in $1+1$, $2+1$ and
$3+1$ dimensions with  $\rho(0)=1$. From now on, we use a constant value of $p_{\rm b} = 0.5$. We note that, although we keep $p_{\rm b}$ constant, the actual
\textit{birth rate} is dependent on $p_{\rm d}$ since the probability of birth is proportional to $p_{\rm b}(1-p_{\rm d})$. To find
the steady state, we examine surviving runs only, looking at an increasing number
of time steps up to $2\times 10^5$ and increments in $p_{\rm d}$ of $5\times 10^{-5}$
as the critical point is approached.

\begin{figure}[tb]
        \centering\noindent
        \includegraphics[width=8cm]{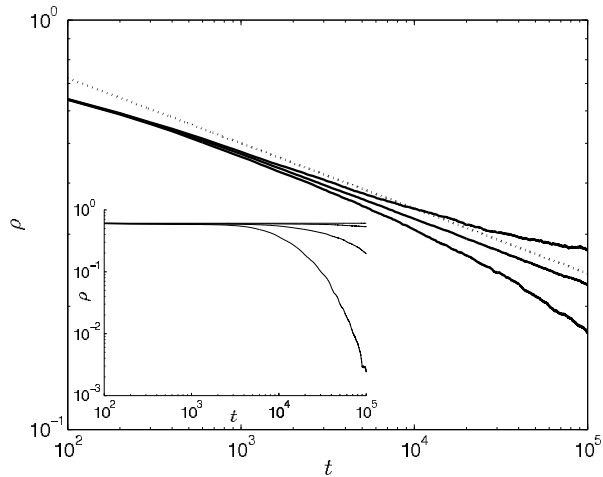}
        \caption{Power-law behaviour for the $1+1$ dimensional model. Solid
        lines represent (from top to bottom) $p_{\rm d}=0.071654$, 0.071754         and 0.071854. The hashed line represents the gradient -0.159             as a guide for the eye. The insert shows non-power law behaviour         for various
        values of $p_{\rm d}$ close to the critical point for the $2+1$ dimensional
        model. The $3+1$ dimensional case is very similar. Information on         how the critical points were found are detailed later in the paper.}
        \label{powerlaw1D}
        \end{figure}

\begin{figure}[tb]
        \centering\noindent
        \begin{tabular}{cc}
        \small{(a)} &  \small{(b)} \\
        \includegraphics[width=7.5cm]{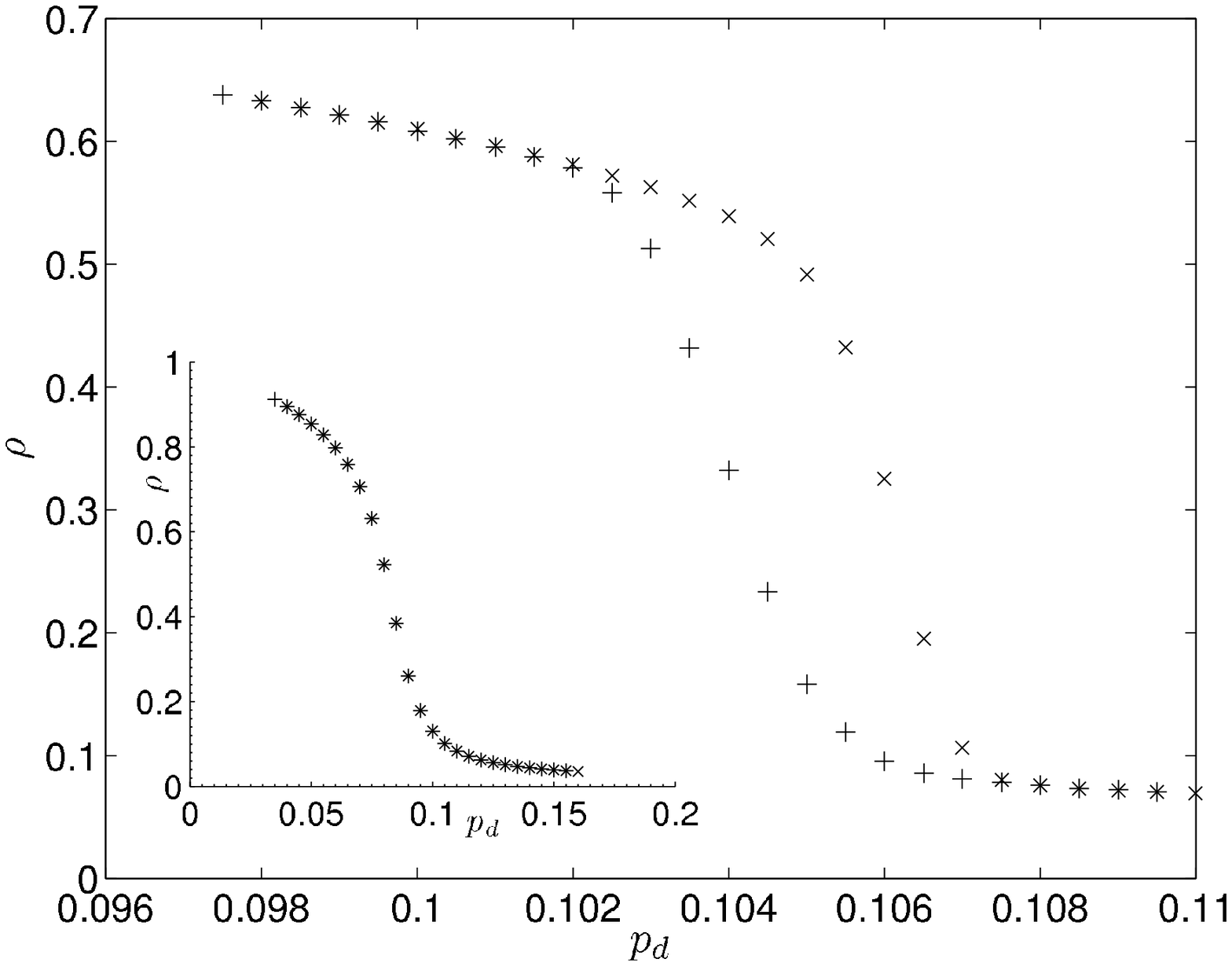} &
        \includegraphics[width=7.5cm]{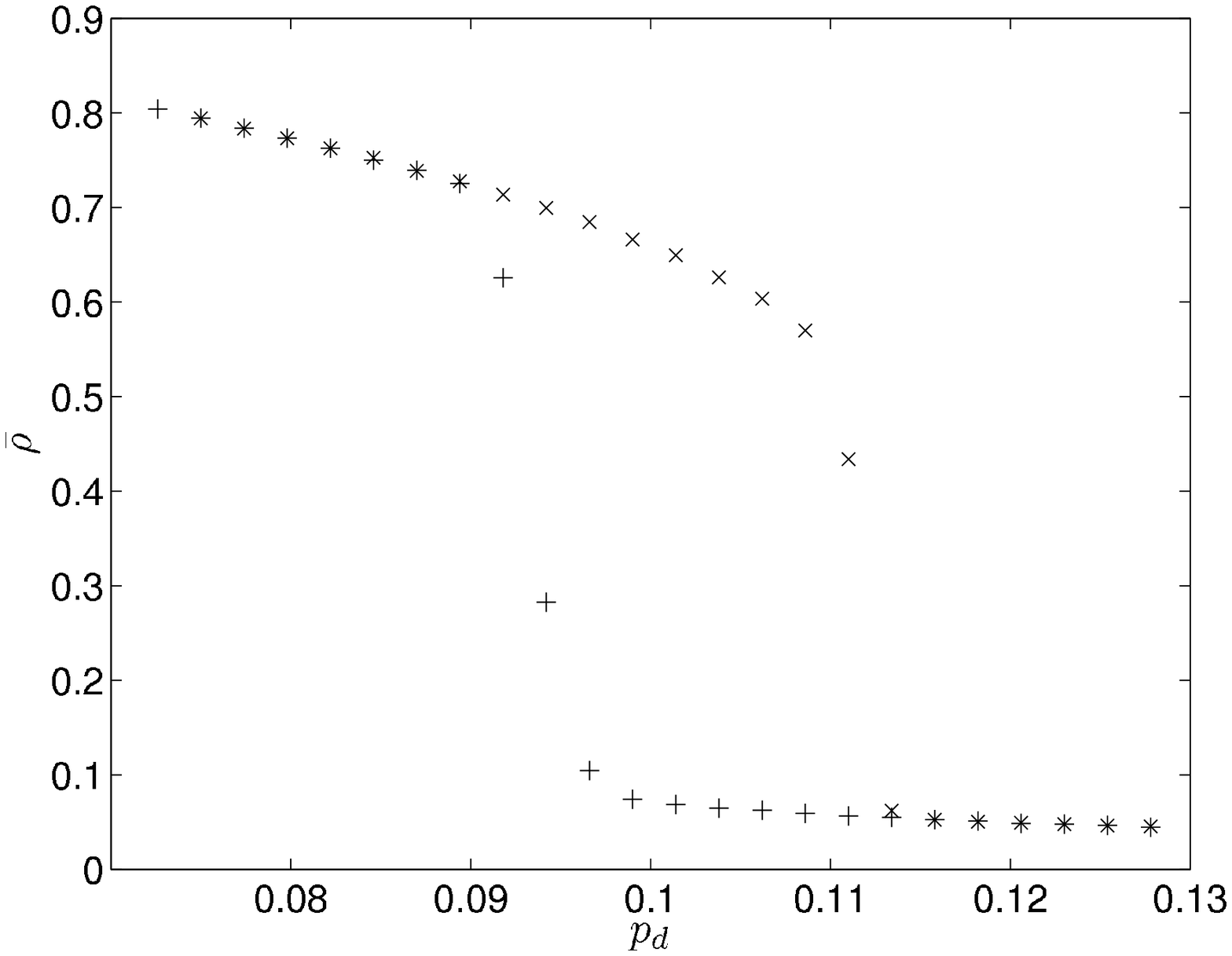}
        \end{tabular}
        \caption{Hysteresis loop for the (a) $2+1$ and (b) $3+1$ dimensional         models
        with $p_{\rm d}$ increasing ($\times$) and decreasing ($+$). The         insert
        in (a) shows no hysteresis occurring in the $1+1$ dimensional model.}
        \label{Hysteresis}
        \end{figure}

We see from the results that we have a strong indication of a first order
phase transition in $2+1$ and $3+1$ dimensions whilst a continuous phase transition
in $1+1$ dimensions. While these results are compelling, we note that since
the simulations were performed on finite lattices and for finite times,
we cannot take them to be conclusive since in such simulations there is always
a non-zero probability of survival for finite $t$ even for $p_{\rm d}>p_{{\rm d}_{\rm c}}$.
Instead we look for power-law behaviour in $\rho(t)$ close to the critical point.
For a continuous phase transition we expect asymptotic power-law behaviour of the order
parameter at the critical point (see \cite{Lubeck_Scaling} for a review)
of the form
        \begin{equation}
        \rho(t) \propto t^{-\delta}.
        \end{equation}
In log-log plots, positive curvature for large $t$ indicates the system is in the active phase
whereas negative curvature implies that the system is in the absorbing phase.
A first order phase transition will therefore be marked by non power-law
behaviour, rather exponential decay of the order parameter for $p_{\rm d} > p_{{\rm d}_{\rm c}}$.
Figure \ref{powerlaw1D} shows the $1+1$ dimensional case with $\rho(t)$ for different
values of $p_{\rm d}$ close to the critical point clearly showing power-law behaviour
at the critical point.
No power-law behaviour is however observed in $2+1$ or $3+1$ dimensions; instead, with the inclusion of spontaneous particle creation at rate $\kappa = 0.005$, hysteresis occurs in both cases as plotted in Figure \ref{Hysteresis},         
which is indicative of first-order phase transitions. However, no hysteresis is observed
in the $1+1$ dimensional case. We examine the first-order
phase transitions in more detail in Section \ref{s: First-order}.

Since the model shows a continuous phase transition in $1+1$ dimensions,
we must ask ourselves to which universality class it belongs. By Grassberger and Janssen's conjecture \cite{Grassberger_On, Janssen_Non}, we would expect it to belong to the universality
class of DP. We investigate this now by looking at
the critical exponents and a scaling function in turn for $1+1$ dimensions only.  

\section{Critical Exponents} \label{s: Critical Exponents}
Finding the critical exponents through steady state simulations is notoriously
difficult due to critical slowing down, finite-size effects, large fluctuations
and the difficulties that arise in finding the critical point. A much more
effective method is that of time-dependent simulations, which has proved to
be a very efficient way of determining the critical exponents and the
critical point for models exhibiting absorbing phase transitions \cite{Grassberger_Torre}. Using this
method, the time-evolution of the model
is observed up to some time $t_{\rm M}$, after beginning with a configuration that is very close to the absorbing state - 2 adjacent particles in this model.
The size of the lattice is made large enough so that the particles never reach the boundary before $t_{\rm M}$. 

We measure the survival
probability $P(t)$, defined as the probability that the system has not reached
the absorbing state at time $t$ and the average number of occupied sites $n(t)$. At $p_{\rm d} = p_{{\rm d}_{\rm c}}$, we expect the following asymptotic power-law behaviour  \cite{Grassberger_Directed}
        \begin{eqnarray}
        P(t) & \propto & t^{-\delta},\\
        n(t) & \propto & t^{\eta}.
        \end{eqnarray}
Away from the critical point, the evolution departs from pure power-law and
so by examining log-log
plots of $P(t)$ and $n(t)$ versus $t$ we can find the critical point by finding
the value of $p_{\rm d}$ that gives a straight line. Generally, however, we expect
corrections to the pure power-law behaviour so that $P(t)$ is more accurately
given as \cite{Grassberger_Directed}
        \begin{equation}  \label{corrections}
        P(t) \propto t^{-\delta}\left(1 + at^{-1}+bt^{-\delta'}+...\right)
        \end{equation}
and similarly for $n(t)$. Here, $\delta '$ represents the correction-to-scaling exponent for $\delta$. More precise estimates for the critical exponents
are obtained by examining the local slope
        \begin{equation}  \label{local slope}
        -\delta(t) = \frac{\ln\left[P(t)/P(t/m)\right]}{\ln(m)}
        \end{equation}
and similarly for $n(t)$, where the critical exponent $\delta$ is given by
$\lim_{t\to\infty}\delta(t)$. Here, $m$ is given as the local range over which
the slope is measured and is typically 5 \cite{Jensen_Critical2,Jensen_Critical3} or 8 \cite{Grassberger_Directed}. Grassberger \cite{Grassberger_Directed}
has shown that
 for the local slope
defined in  (\ref{local slope}), we have the following behaviour
        \begin{equation}
        \delta(t) = \delta + at^{-1} + b\delta't^{-\delta'}+...
        \end{equation}
and again, similarly for $n(t)$. Thus if we plot $\delta(t)$ versus $t^{-1}$,
we have that the critical exponent $\delta$ is given by the intercept with
the $y$ axis and any curvature would indicate a correction-to-scaling
exponent less than 1.

\begin{figure}[tb]
        \centering\noindent
        \begin{tabular}{cc}
        \small{(a)} &  \small{(b)} \\
        \includegraphics[width=7.5cm]{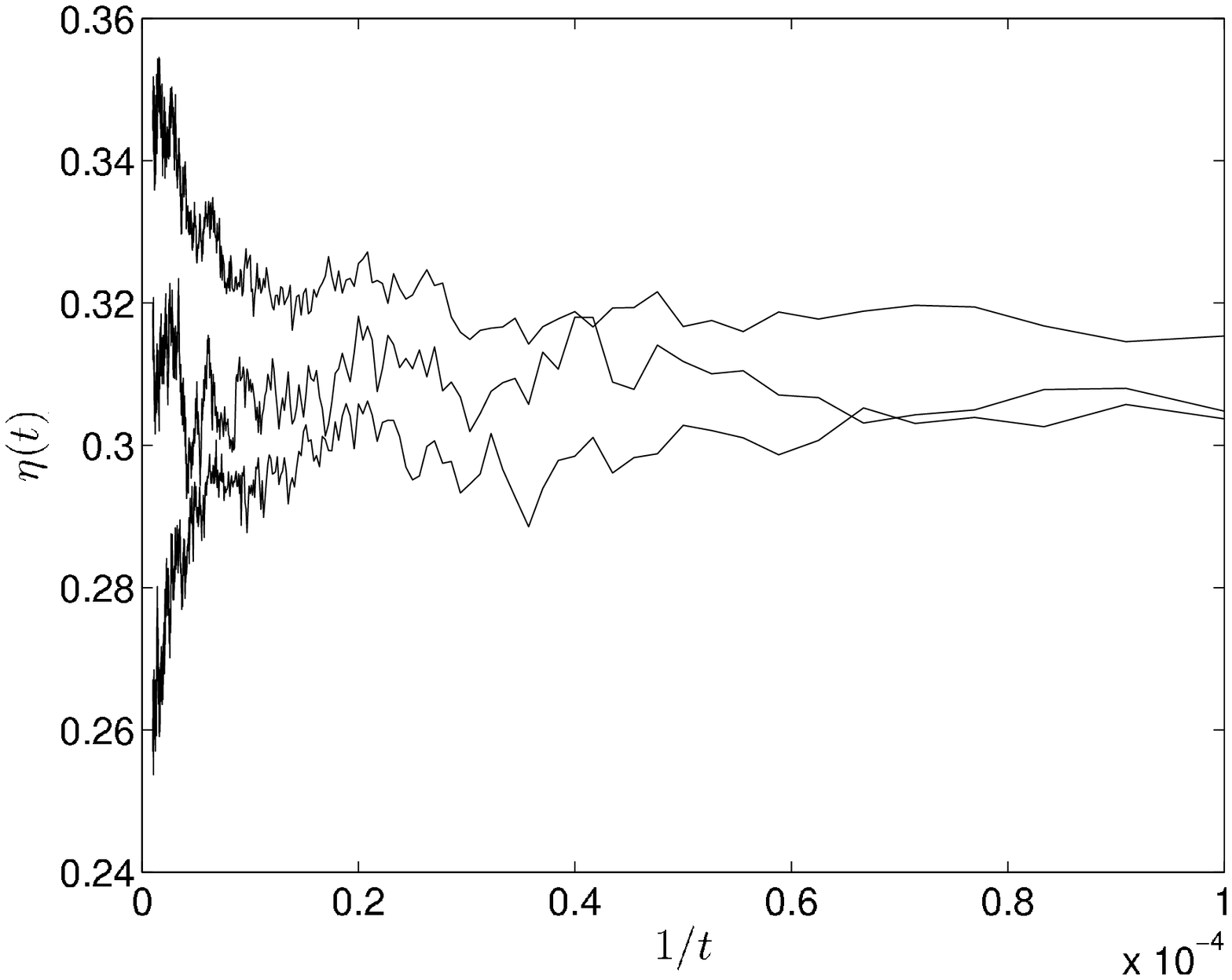} &
        \includegraphics[width=7.5cm]{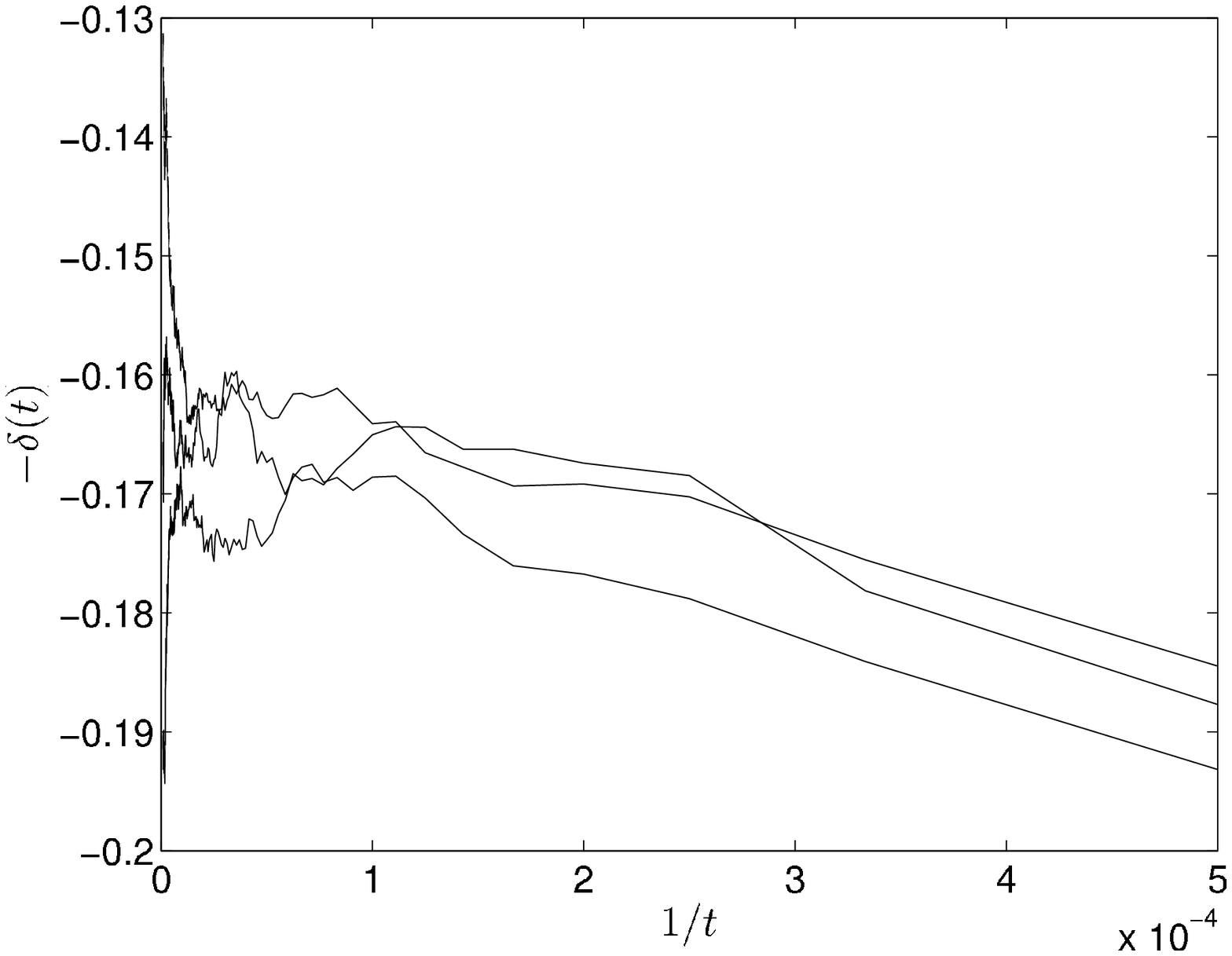}
        \end{tabular}          
        \caption{Plots of (a) $\eta(t)$ and (b) $-\delta(t)$ up to $t = 10^6$.         From top to bottom, $p_{\rm d} = 0.071746$, 0.071754 and 0.071762.}
        \label{Critical Point}
        \end{figure}

\begin{figure}[tb]
        \centering\noindent
        \begin{tabular}{cc}
        \small{(a)} &  \small{(b)} \\
        \includegraphics[width=7.5cm]{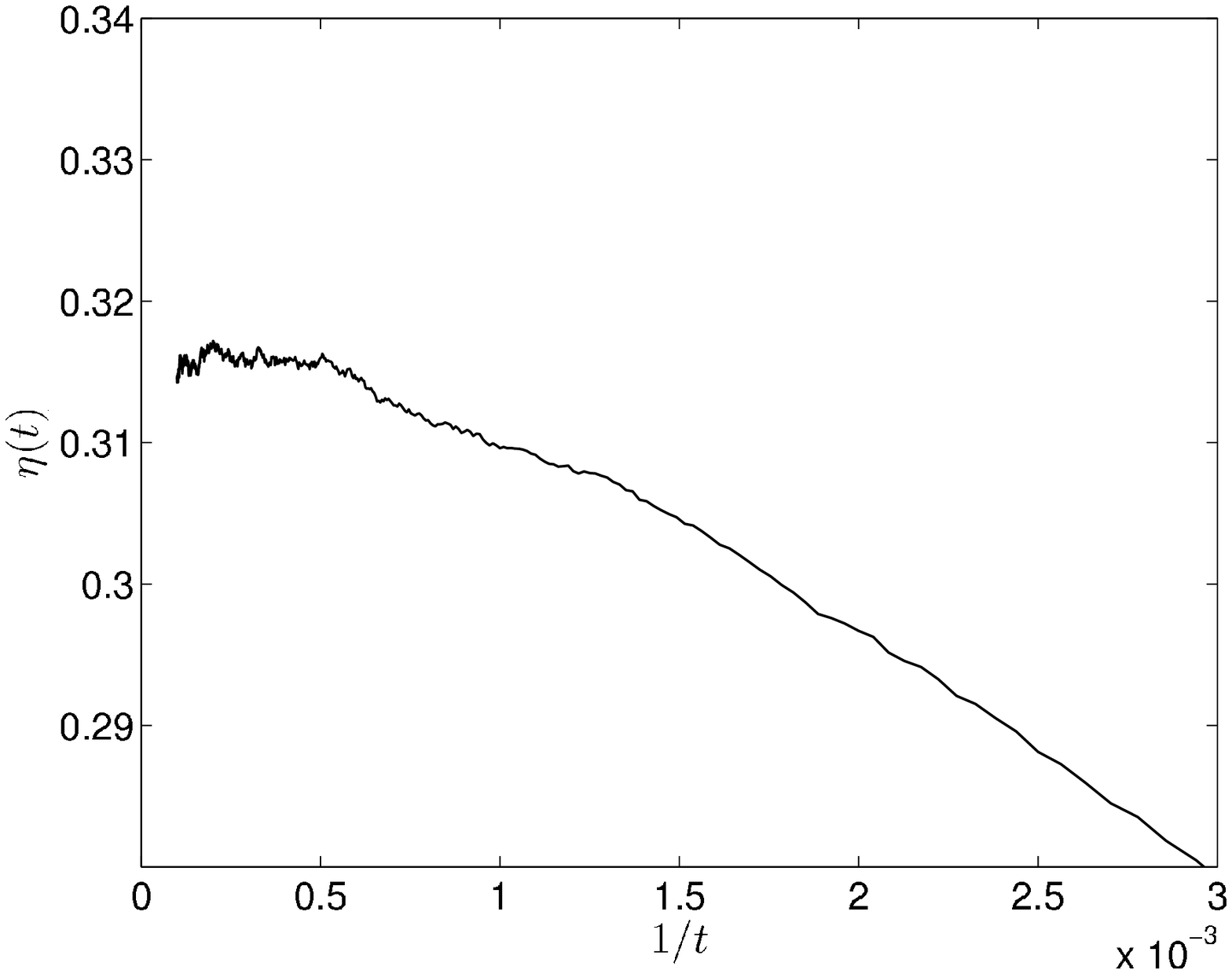} & 
        \includegraphics[width=7.5cm]{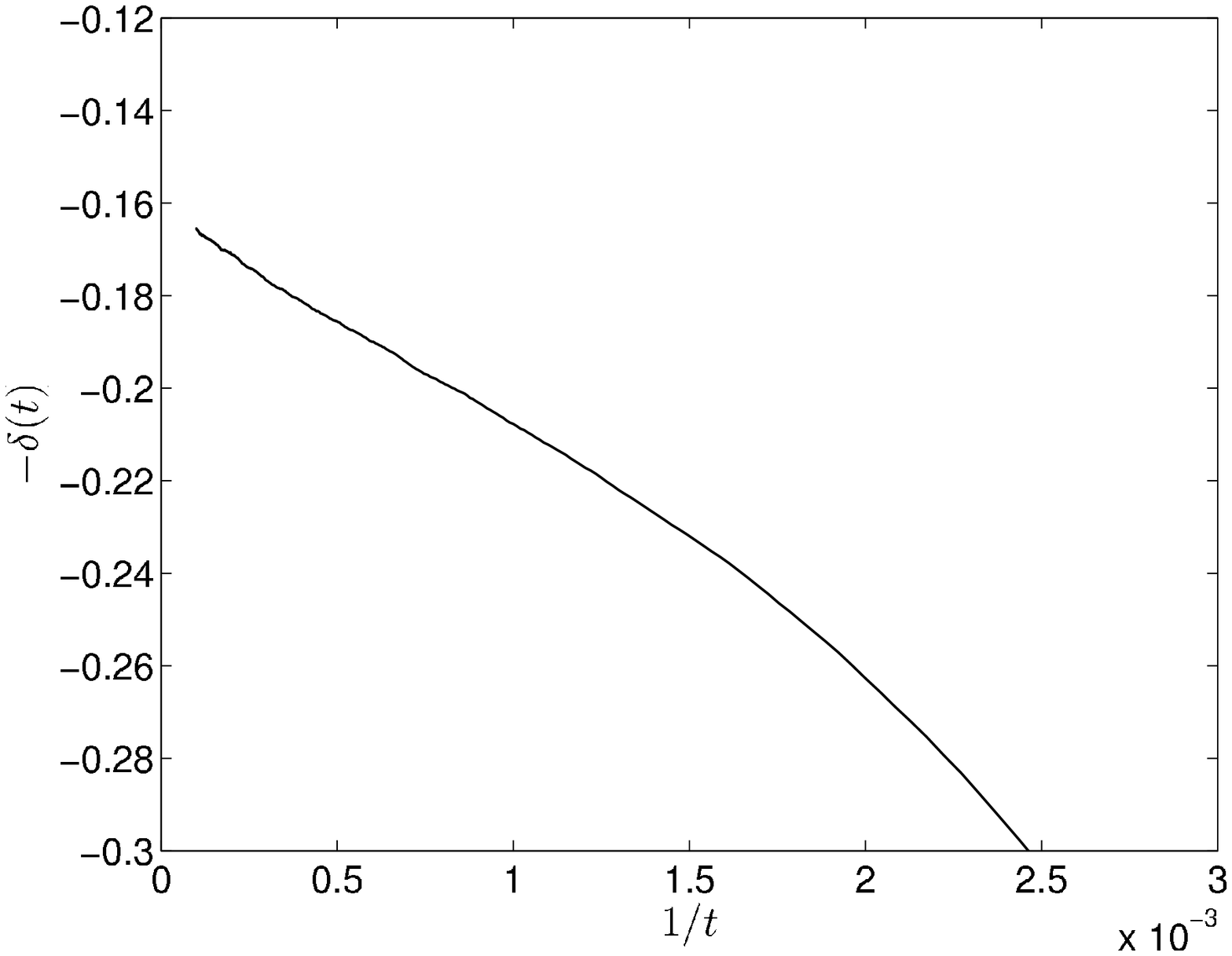}
        \end{tabular}
        \caption{Plots of (a) $\eta(t)$ and (b) $-\delta(t)$ up to $t=10^4$
        at $p_{\rm d} = p_{{\rm d}_{\rm c}} = 0.071754$.}
        \label{Exponents}
        \end{figure}

We plot in Figure \ref{Critical Point}, $\eta(t)$ and $\delta(t)$ for up to
$t=10^6$ and over $10^5$ runs.
We find that the data is in fact very noisy, leading to inaccurate results
for the critical exponents. However,
it is clear that the gradient increases for $p_{\rm d} = 0.071746$ and decreases for
$p_{\rm d} = 0.071762$ for large $t$ whereas we have an approximately straight line for $p_{\rm d} =
0.071754$, especially for $\eta(t)$. This therefore gives a value of $p_{{\rm d}_{\rm c}}
= 0.071754 \pm 0.000004$. We plot in Figure \ref{Exponents} $\eta(t)$ and         
$\delta(t)$ for $p_{\rm d} = 0.071754$ but this time up to $t = 10^4$ and over
$2.5 \times10^6$ runs to improve the accuracy of the results. From this
plot we can read off the values $\eta = 0.312\pm0.002$ and $\delta = 0.160\pm0.002$,
which are in agreement with the best currently known DP values of $\eta = 0.313686$ and $\delta = 0.159464$ \cite{Jensen_Low}.

\section{Scaling Functions} \label{s: Scaling Functions}
For models to belong
to the same universality class, both the critical exponents and the scaling
functions must be the same. Having shown that the critical exponents $\delta$
and $\eta$ for our model are in agreement with the DP values, we
turn our attention now to the scaling function for the
probability of survival, $P(\Delta,L,t)$, where $\Delta = p_{{\rm d}_{\rm c}}-p_{\rm d}$. We have the scaling ansatz \cite{Lubeck_Universal}
        \begin{equation} \label{ScalingFn}
        P(\Delta, L, t) \sim \lambda^{-\beta}R_1(\Delta\lambda,
          L\lambda^{-\nu_\perp},t\lambda^{-\nu_\parallel})
        \end{equation}
for any $\lambda > 0$. $\beta$ is the critical
exponent associated with the survival probability according to 
        \begin{equation}
        P \sim \Delta^{\beta}
        \end{equation}
as $t\longrightarrow\infty$ and $\nu_\parallel$ and $\nu_\perp$ are the critical
exponents associated with the temporal and spatial correlation lengths respectively.
As in the previous section, we begin our simulations from a single
seed with $L$ large enough so that we can ignore the $L$-dependence in
 (\ref{ScalingFn}) and examine
        \begin{equation}
        P(\Delta, t) \sim \lambda^{-\beta}R_2(\Delta\lambda,
          t\lambda^{-\nu_\parallel}),
        \end{equation}
only. The functions $R_i$ are non-universal, i.e. they are unique to each
model. However, we may introduce metric factors $a_\Delta$ and $a_t$,
        \begin{equation}
        P(\Delta, t) \sim \lambda^{-\beta}\bar R(a_\Delta\Delta\lambda,
          a_t t\lambda^{-\nu_\parallel})
        \end{equation}
so that $\bar R$ is a universal scaling function. Now all of non-universal,
system-dependent features, such as the update scheme, boundary conditions, lattice structure, etc. are contained in these
non-universal metric factors.
 
If we choose $\lambda = (a_t t)^{1/\nu_\parallel}$,
then, since $\delta = \beta/\nu_\parallel$,
        \begin{equation} \label{DataCollapse}
        P(\Delta, t) \sim (a_t t)^{-\delta} \bar R(a_\Delta\Delta(a_t            t)^{1/\nu_\parallel},1)
        \end{equation}
so we would expect to observe a data collapse by plotting $(a_t t)^{\delta}P$ versus
$a_\Delta\Delta(a_t t)^{1/\nu_\parallel}$ for all models belonging to the
same universality class.
To find the metric factors $a_t$ and $a_\Delta$, we use the normalisations
        \begin{equation}
        \bar R(1,\infty) = \bar R(0,1) = 1.
        \end{equation}
Then, choosing $\Delta = 0$ and $\lambda = (a_t t)^{1/\nu_\parallel}$ we
have that $a_t$ is given by the amplitude of 
        \begin{equation}
        P \sim (a_t t)^{-\delta}.
        \end{equation}
Similarly, setting $\lambda = (a_\Delta\Delta)^{-1}$ and letting $t\longrightarrow\infty$,
$a_\Delta$ is given by the  amplitude of
        \begin{equation}
        P \sim (a_\Delta\Delta)^{\beta},
        \end{equation}
where, since $\lambda > 0$, we must have $\Delta > 0$.

We now compare the scaling function for our model with that for \textit{directed
bond percolation} which belongs to DP. The model begins with an initial number of active sites and proceeds in time with site $i$ becoming active at time $t+1$ if site $i+1$ and/or site $i-1$ is active at time $t$ and there exists a bond (with probability $p$) between this active site and the site $i$.
The best estimate of the critical value of $p$ is $p_{\rm c} = 0.644700185(5)$
\cite{Jensen_LowIII} where, for $p \ge p_{\rm c}$, the active sites percolate
the infinite system.   

\begin{figure}[tb]
        \centering\noindent
        \includegraphics[width=8cm]{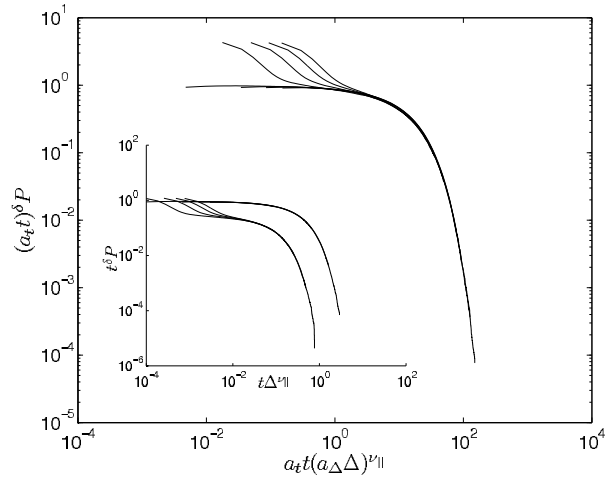}
        \caption{Data collapse for  both the directed bond percolation model         and our own using metric factors. Insert: Data collapse for both         the directed bond percolation  model (top right) and our own (bottom         left).}
        \label{DataCollapses}
        \end{figure}

As was the case in our model, we begin with a single seed - just one active site at the origin here - 
and record the probability of survival $P_{DP}(t)$, i.e. the probability
that there is at least one active site at time $t$. We denote the metric
factors with a subscript $DP$ for the \textit{directed bond percolation} model and
subscript $P$ for our model. Excluding the metric factors, we
plot in the insert in Figure \ref{DataCollapses} the two separate data collapses for $P$ for both the directed bond percolation
model and for our own by plotting $Pt^{\delta}$ vs. $t\Delta^{\nu_\parallel}$.
We use the best known DP values of $\delta = 0.159464$ and $\nu_\parallel = 1.733847$ \cite{Jensen_Low}
and see clear data collapses in both models, especially for large $t$. Using now the obtained
values for the metric factors, $a_{t,DP} = 1.57$, $a_{\Delta, DP} = 7.51$,
$a_{t,P} = 3137.30$ and $a_{\Delta, P} = 0.2$,
we have a complete data collapse of all the data from both models as shown
in the main plot in Figure \ref{DataCollapses}, indicating that the scaling function for both models is identical. 

Having examined both the critical exponents and the scaling functions, we
conclude that our model belongs to DP in $1+1$ dimensions.

\section{First-Order Phase Transitions} \label{s: First-order}
Due to DP having a critical dimension, $d_{\rm c} = 4$ \cite{Cardy_Directed}, it is surprising that this model belongs to DP in $1+1$ dimensions only due to the first-order transition in higher dimensions. This is not a unique property to this model. A stochastic cellular automaton  model developed by Bidaux, Boccara and Chat\'e (BBC) \cite{Bidaux} is known to belong to DP in $1+1$ dimensions \cite{Jensen_Universality} yet also displays a first-order transition in higher dimensions. 

In order to examine the first-order phase transitions, we borrow a technique developed by J. Lee and J.M. Kosterlitz
\cite{Lee_New}, which allows one to determine the order of a phase transition
in an equilibrium system. The method detects a temperature-driven first-order phase transition by Monte Carlo simulations in a finite
system of volume $L^{\rm d}$ with periodic boundary conditions by examining the
histogram of the energy distribution
        \begin{equation}
        N(E; \beta, L) = \mathcal{N}Z^{-1}(\beta, L)\Omega(E,L)\exp(-\beta
        E).
        \end{equation}
$\mathcal{N}$ is the number of MC sweeps, $Z$ is the partition function
and $\Omega$ is the number of states with energy $E$. For the $q$-state Potts
models with $q$ ordered and one disordered state, $N$ has a characteristic
double-peak close to $T=T_c$ for energy values $E_1(L)$ and $E_2(L)$ corresponding
to the ordered and disordered states. The peaks are separated by a minimum at $E_{\rm m}(L)$. If we define
        \begin{equation}
        A(E; \beta,L, \mathcal{N}) = -\ln N(E; \beta,L, \mathcal{N}),
        \end{equation}
then at $\beta = \beta_c(L)$ defined by $A(E_1; \beta,
L) = A(E_2; \beta, L)$,
        \begin{equation}
        A(E_m; \beta,L, \mathcal{N}) - A(E_1; \beta,L, \mathcal{N}) = \Delta
        F 
        \end{equation}
where $\Delta F$ is the bulk free-energy barrier between the states. At a
first-order phase transition, $\Delta F \sim L^{{\rm d}-1}$ for $L\gg \xi$, where
$\xi$ is the correlation length, whereas $\Delta F$
is independent of $L$ at a continuous phase transition.

\begin{figure}[tb]
        \centering\noindent
        \includegraphics[width=8cm]{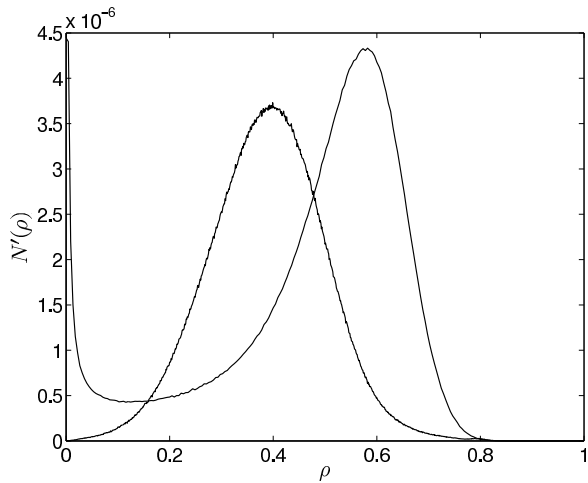}
        \caption{Normalised histogram, $N'(\rho)$ for different population         densities in the $1+1$ dimensional (left) and $2+1$ (right) dimensional         cases showing the results for a continuous and first-order phase         transition respectively.}
        \label{Histogram}
        \end{figure}

Due to our model being out of equilibrium, we continue only in the spirit
of the above method and examine the histogram for the population density
$N(\rho)$. For a first-order phase transition, we expect a double-peaked
structure at $\rho_0$ and at $\rho_+ = \rho_- = 0.5$ due to the corresponding phase
coexistence. For $p_{\rm d} > p_{{\rm d}_{\rm c}}$ we expect $N(\rho_0) > N(\rho_+)$ due to
the greater chance of extinction, and likewise
for $p_{\rm d} < p_{{\rm d}_{\rm c}}$ we expect $N(\rho_0) < N(\rho_+)$ due to the greater
chance of survival. This in fact gives
us an excellent method for determining the critical point since it will be
marked by $N(\rho_0)$ and $N(\rho_+)$ being equal. At a continuous phase transition, however, due to the power-law behaviour $\rho \sim t^{-\delta}$,
we expect $N(\rho_0)=0$ at the critical point. Both expectations are confirmed in Figure \ref{Histogram},
where we plot the histogram at the critical points for both the $1+1$ and $2+1$
dimensional cases.  

\begin{figure}[tb]
        \centering\noindent
        \includegraphics[width=8cm]{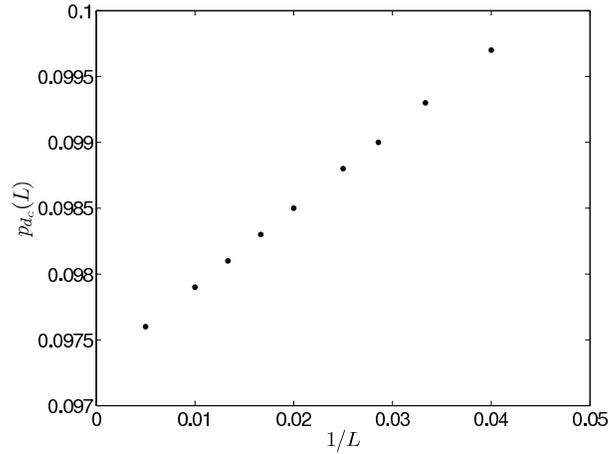}
        \caption{Plot to obtain an approximation for the critical point in
        the $2+1$ dimensional model by extrapolation of the points to the         intersection with the $y$-axis.}
        \label{CriticalPoint2D}
        \end{figure}

As has been mentioned, the phase-coexistence enables the critical value $p_{{\rm d}_{\rm c}}$ to be estimated. Iteratively finding the critical value for large enough $t$ and different values
of $L$ and then plotting $p_{{\rm d}_{\rm c}}(L)$ vs. $1/L$ in Fig. \ref{CriticalPoint2D}, we are able to obtain, by an extrapolation to the intercept with the $y$
axis, the value of the critical point in $2+1$ dimensions
to be $0.0973\pm 0.0001$. Unfortunately, no conclusive numerical evidence
has yet been obtained in $3+1$ dimensions to give an accurate value for $p_{{\rm d}_{\rm c}}$.

\section{Concluding Remarks}  \label{s: Conclusions}
By examining the critical exponents $\delta$ and $\eta$ and the scaling function for the survival probability, we have shown that the $1+1$ dimensional
version of our model belongs to the class of DP. Higher dimensions however
do not, surprisingly, belong to this class due to the observed first-order
phase transitions. In order to examine these transitions we used a technique
inspired by equilibrium systems to find the value of the critical point in
$2+1$ dimensions.
\\ \\
In order to answer the question about why our model does not belong to DP
in all dimensions, we must look at how our model differs from others
in the class. One difference to other models such as the \textit{contact
process} \cite{Harris} or the \textit{Domany-Kinzel} \cite{Kinzel, Domany}  model is that, here, population growth requires two particles meeting, whereas death
involves one particle only. This requirement results in the existence of the
critical population density, meaning that extinction can occur either because
the death rate is too large and/or because the population density is too
small. The previously mentioned BBC model shares this feature since they
too observed a critical value of the initial concentration of ``alive'' sites
below which the system decayed to a zero concentration state. With this inclusion, it is remarkable that the $1+1$ dimensional
version of our model belongs to DP at all and only goes to highlight the robust nature of the universality class. 
 
It seems likely that the unusual behaviour of both our model and that of BBC displaying DP behaviour in $1+1$ dimensions, whilst first-order transitions in higher dimensions, is due to the critical population density/concentration
present in both models. In our model, with the given rules for growth and decay, it is clear
that the population will become more dependent on the density as the
dimensionality of the system increases since particles will find it progressively
harder
to meet one another before they die. Time-dependent simulations carried out in higher dimensions, for example,
needed increasingly large populations at $t=0$ for \emph{initial} population growth as $p_{\rm d}$
was increased, whereas two particles were sufficient in $1+1$ dimensions even
for $0<p_{\rm d}-p_{{\rm d}_{\rm c}}\ll 1$. 

To examine whether the first-order transitions are a result of the critical
population density, we changed our model to include single-particle reproduction, $A \longrightarrow 2A$, at rate $c$. From mean field analysis, this inclusion
eradicates the critical population density and initial numerical analysis
appears to show that this changes the transition to continuous in at least the $2+1$ dimensional case. 

In the $1+1$ dimensional case, not only is it easier for
particles to meet, but the larger fluctuations in the population are likely
to be enough to induce the observed continuous phase transition as has been
known to happen \cite{Marro}. This was tested by introducing two separate types
of particles into the system, $L$ and $R$ particles, where the $L$ particles could only move left and $R$ particles to the right. All other rules remained
the same except that both an $L$ and a $R$ particle were required for reproduction.
This modification to our model reduced the size of the fluctuations in the
overall population and, again, early numerical simulations appear to show
that this results in the transition changing to first-order.

Clearly, more work will however have to be carried out to confirm the above observations
and to find out exactly what is happening at the phase transition.

\ack

We would like to thank Daniel Lawson and Gunnar Pruessnar
for useful discussions and Uwe T\"auber and Iwan Jensen for helpful email correspondence. Alastair Windus would also like to thank the Engineering
and Physical Sciences Research Council (EPSRC) for his Ph.D. studentship. 
\section*{References}

\bibliography{bibliography}

\end{document}